\documentclass[sigconf]{acmart}

\usepackage{tabularx}
\usepackage{multirow}
\usepackage{multicol}
\usepackage{ulem}
\usepackage{booktabs}
\newcolumntype{C}{>{\centering\arraybackslash}X}
\usepackage{algorithm}
\usepackage{algorithmic}
\usepackage{newfloat}
\usepackage{listings}
\usepackage{float}
\usepackage{hyperref}
\usepackage[table,xcdraw]{xcolor}
\usepackage{balance}

\AtBeginDocument{%
  }

\makeatletter
\def\@fnsymbol#1{\ensuremath{\ifcase#1\or \dagger\or *\or \ddagger
   \or \mathsection\or \mathparagraph\or \|\or **\or \dagger\dagger
   \or \ddagger\ddagger \else\@ctrerr\fi}}
\makeatother
\setcopyright{acmlicensed}
\copyrightyear{2025}
\acmYear{2025}
\setcopyright{acmlicensed}\acmConference[CIKM '25]{Proceedings of the 34th ACM International Conference on Information and Knowledge Management}{November 10--14, 2025}{Seoul, Republic of Korea}
\acmBooktitle{Proceedings of the 34th ACM International Conference on Information and Knowledge Management (CIKM '25), November 10--14, 2025, Seoul, Republic of Korea}

\begin{document}

\title{Augmenting Limited and Biased RCTs through Pseudo-Sample Matching-Based Observational Data Fusion Method}
\author{Kairong Han}
\orcid{0000-0003-0003-2312}
\affiliation{%
  \institution{Zhejiang University}
  \city{Hangzhou}
  \country{China}
}
\email{zju_handso@163.com}

\author{Weidong Huang }
\orcid{0009-0004-6704-5323}
\affiliation{%
  \institution{Didi Chuxing}
  \state{Beijing}
  \country{China}
}
\email{huangweidong@didiglobal.com}

\author{Taiyang Zhou}
\orcid{0009-0005-1413-6771}
\affiliation{%
  \institution{Didi Chuxing}
  \state{Beijing}
  \country{China}
}
\email{zhoutaiyang@didiglobal.com}

\author{Peng Zhen}
\orcid{0009-0008-5062-3798}
\affiliation{%
  \institution{Didi Chuxing}
  \state{Beijing}
  \country{China}
}
\email{zhenpeng@didiglobal.com}

\author{Kun Kuang}
\authornote{Corresponding author.}
\orcid{0000-0001-7024-9790}
\affiliation{%
  \institution{Zhejiang University}
  \city{Hangzhou}
  \country{China}
}
\email{kunkuang@zju.edu.cn}

\renewcommand{\shortauthors}{Kairong Han, Weidong Huang, Taiyang Zhou, Peng Zhen, \& Kun Kuang}
\begin{abstract}
In the online ride-hailing pricing context, companies often conduct randomized controlled trials (RCTs) and utilize uplift models to assess the effect of discounts on customer orders, which substantially influences competitive market outcomes.        However, due to the high cost of RCTs, the proportion of trial data relative to observational data is small, which only accounts for 0.65\% of total traffic in our context, resulting in significant bias when generalizing to the broader user base.        Additionally, the complexity of industrial processes reduces the quality of RCT data, which is often subject to heterogeneity from potential interference and selection bias, making it difficult to correct.        Moreover, existing data fusion methods are challenging to implement effectively in complex industrial settings due to the high dimensionality of features and the strict assumptions that are hard to verify with real-world data.        To address these issues, we propose an empirical data fusion method called pseudo-sample matching.        By generating pseudo-samples from biased, low-quality RCT data and matching them with the most similar samples from large-scale observational data, the method expands the RCT dataset while mitigating its heterogeneity.        We validated the method through simulation experiments, conducted offline and online tests using real-world data. In a week-long online experiment, we achieved a 0.41\% improvement in profit, which is a considerable gain when scaled to industrial scenarios with hundreds of millions in revenue. In addition, we discuss the harm to model training, offline evaluation, and online economic benefits when the RCT data quality is not high, and emphasize the importance of improving RCT data quality in industrial scenarios.  Further details of the simulation experiments can be found in the GitHub repository  \href{https://github.com/Kairong-Han/Pseudo-Matching}{here}.

\end{abstract}




\maketitle

\section{Introduction}

Ride-hailing, such as Uber~\cite{calo2017taking} and DiDi, is a great example of the creative economy in which owners of private vehicles provide a convenient service to the public for a modest fee~\cite{CONTRERAS201863}, which has transformed the way people travel. A well-designed pricing strategy can enhance user loyalty, increase the user base, improve the customer experience, and maximize revenue while minimizing costs~\cite{yan2020dynamic}.   To determine optimal pricing strategies, ride-hailing companies conduct Randomized Controlled Trials (RCTs)~\cite{stolberg2004randomized}, also known as A/B tests.   Specifically, when users request a ride, companies randomly apply discounts from a set of alternatives and observe user behavior, allowing them to calculate the heterogeneous causal effect of discounts on order rates.   This information is crucial for selecting discounts, budget control, cost optimization, and profit calculation.

However, the real-world ride-hailing pricing scenario is highly complex, and the idealized process described above may not perform well in practical industrial applications. Specifically, companies encounter two core challenges:

 \textbf{(1) The generalizability of a small amount of RCT data is limited.} The target customer base is vast, and the feature space is highly dimensional~\cite{hu2023customer}. And the cost of conducting large-scale RCT experiments is also prohibitively high. In our scenario, the RCT data represents only about 0.65\% of the total traffic. Confounding factors such as different scenarios, weather conditions, passenger demographics, and supply-demand relationships all influence the causal effect of discounts on order rates. Consequently, when applied to real-world outcomes, the uplift model trained on this small portion of RCT data exhibits significant bias.

\begin{figure}
    \centering
    \includegraphics[width=0.9\linewidth]{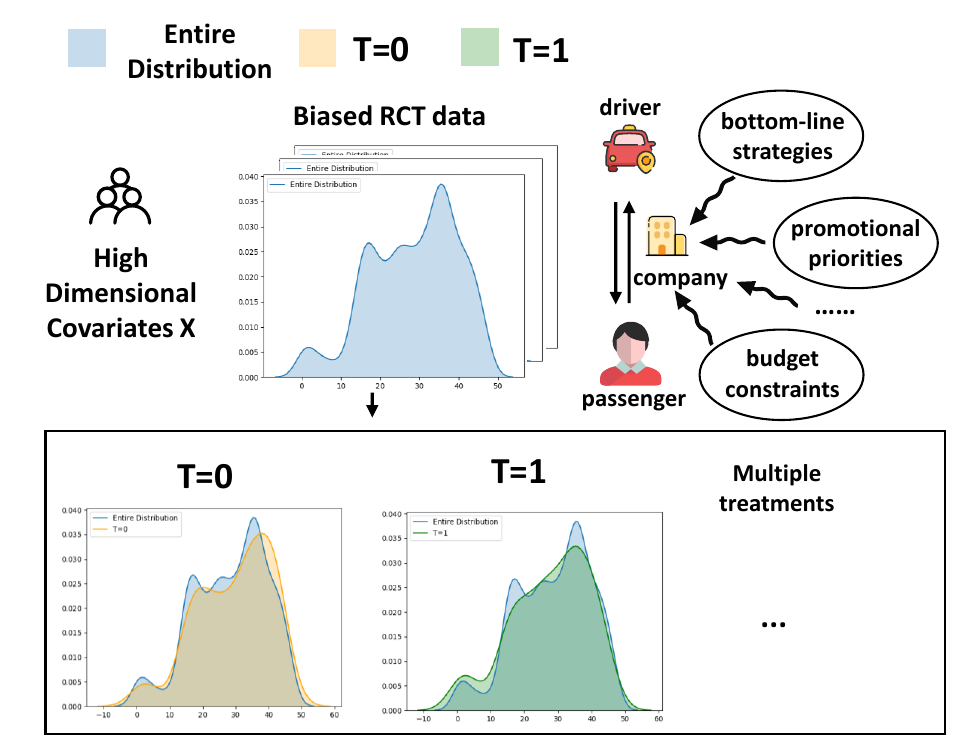}
    \caption{Heterogeneity diagram from real-world data. Complex link interference and potential selection bias lead to biased RCTs. The user's high-dimensional covariate $X$ has different distributions across the different treatment groups $T$ in the biased RCT.}\label{fig:induction_yzx}
    \vspace{-2em}
\end{figure}

 \textbf{(2) Complex relationships introduce inherent heterogeneity in randomized controlled trials.} In real-world scenarios, the experimental connections are intricate, making idealized random experiments impractical, as shown in Figure \ref{fig:induction_yzx}.  When conducting RCTs, heterogeneity in the data can arise from several factors, including budget constraints, promotional priorities in specific cities, implementation errors in complex engineering systems, bottom-line strategies, potential changes in population characteristics over time, and even interdepartmental influences within the company.  These potential interferences and selection biases result in groups in RCTs that are not entirely random, leading to discrepancies in data distribution across different treatment groups. Consequently, the inherent heterogeneity results in significant deviations in the final trained uplift model.

To address these challenges, data fusion~\cite{colnet2024causal} methods are employed to incorporate biased but more comprehensive observational data on population characteristics, thereby enhancing the generalization of RCT data.  However, current data fusion techniques face three significant challenges in industrial scenarios: (1) \textbf{The assumptions are often overly stringent and difficult to verify with real-world, high-dimensional data.}  For instance, most existing methods require positive assumptions and Stable Unit Treatment Value (SUTV) assumptions~\cite{yao2021survey,zhao2024learning}, which are challenging to validate in practical settings.   (2) \textbf{Traditional approaches treat RCTs as the gold standard.}  However, RCT data may exhibit certain biases and heterogeneity due to potential selection bias and uncontrollable interference in complex industrial scenarios.  Overlooking these biases may amplify estimation errors in downstream tasks, ultimately affecting the conclusions. (3) \textbf{Some depth-based data fusion algorithms are prohibitively expensive and complex.}  For example, the  CorNet~\cite{hatt2022combining} necessitates training an additional correction model for each intervention group.  However, in industrial scenarios, there may be many treatment groups. For example, there are 33 discrete treatments in our ride-hailing. Training a separate correction model for each intervention group is too expensive and difficult to iterate. Moreover, the high dimensionality of the data further poses challenges to existing methods.

Therefore, we propose a data fusion method based on pseudo-sample matching. It simply matches similar samples in observational data by constructing pseudo-samples of biased RCT data, thereby enhancing the effect of the uplift model. We first proved the effectiveness of the method in simulation experiments and then applied the method to offline and online tests on real-world data, which achieved positive results in online experiments. Our contributions are summarized as follows:

\begin{itemize}
    \item To address the issues of potential heterogeneity and significant generalization bias due to insufficient sample sizes in industrial RCTs, we propose a method for combining RCT and observational data based on pseudo-sample matching.
    \item We validated and analyzed the effectiveness of the proposed method through simulation experiments, offline and online experiments. In a week-long online experiment, we achieved a 0.41\% improvement in profit, which is a considerable gain when scaled to industrial scenarios with hundreds of millions in revenue.
    \item We discuss the biases present in RCT data within industrial settings and their potential risks. Our method is grounded in real-world scenarios, offering strong practical applicability.  With flexible sample bucket and fusion ratio selection, the method can be easily and effectively integrated into a wide range of industrial applications.
\end{itemize}

\section{Related Work}

\subsection{Uplift Model}

Uplift modeling~\cite{zhang2021unified, devriendt2018literature, olaya2020survey} is a predictive technique that estimates the incremental impact of a treatment (e.g., a marketing action) on individual behavior, which is widely used in recommendation systems\cite{bobadilla2013recommender,fu2023end,fu2025forward}. It is closely related to the estimation of heterogeneous treatment effects in causal inference~\cite{xie2012estimating, pearl2010causal,zhao2024networked,wu2024causality}. A key distinction, however, lies in their assumptions: uplift modeling methods are typically designed for data from randomized experiments and often omit explicit assumptions. In contrast, methods from the treatment effect heterogeneity literature specify assumptions. ~\cite{zhang2021unified}. Many machine learning methods have been adapted for estimating heterogeneous treatment effects, such as random forest-based methods~\cite{wager2018estimation}, Bayesian algorithms~\cite{alaa2017bayesian,zhang2020learning}, and deep learning algorithms~\cite{johansson2016learning,yoon2018ganite,yang2025leveraging}. To evaluate the performance of the uplift model, when the Ground-truth causal effect is known, the researchers use the PEHE~\cite{hill2011bayesian} and MAPE ~\cite{de_Myttenaere_2016} metrics. When the Ground-truth causal effect is unknown, the researchers construct the Qini curve by ranking the sample estimated causal effect and then obtaining the Qini coefficient for quantitative evaluation.

\subsection{Combining Randomized Trials and Observational Studies}

Although RCT is the golden standard of causal inference and the uplift model~\cite{cartwright2007rcts}, the cost of RCT is too high, so it is difficult to obtain a large amount of RCT data, resulting in a small amount of RCT data and an inability to generalize well to the target population~\cite{colnet2024causal,chen2024learning}. Therefore, some studies focus on combining RCT and observational data, either to ensure the unconfoundedness of the observational analysis or to improve (conditional) average treatment effect estimation (ATE). When observational data have no treatment and outcome information, to identify the ATE on the target population, most existing methods 
rely on direct modeling of the selection score previously introduced. The selection score adjustment methods include IPSW~\cite{cole2010generalizing,stuart2011use,buchanan2018generalizing}, stratification~\cite{tipton2013improving,o2014generalizing}, and leverage regression formulation. The IPSW estimator of the ATE is defined as the weighted difference of average outcomes between the treated and control group in the trial, to account for the shift of the covariate distribution from the RCT sample to the target population. Stratification is proposed as a solution to mitigate the risks of extreme weights in the IPSW formula. Regression formulation, known as plug-in g-formula estimators~\cite{dahabreh2020extending}, fits a model of the conditional outcome mean among trial participants, rather than modeling the probability of trial participation. When observational data have treatment and outcome information, efficiency improvements can be obtained~\cite{huang2021leveraging}, and we could deal with unmeasured confounders in
observational data. One way is using an assumption on secondary outcomes or surrogates~\cite{athey2020combining}. The other way is using the bias function to de-bias confounding. For instance, Kallus et al.~\cite{kallus2018confounding} use observational data to estimate a flexible, but biased function for the heterogeneous
treatment effect and then aim to remove the bias using the randomized data. Cheng et al.~\cite{cheng2021adaptive} estimate two separate estimators: one biased estimator on observational data, and another unbiased on randomized data. Hatt et al.~\cite{hatt2022combining} introduce a two-step framework. First, they use observational data to learn a shared structure, and then use randomized data to learn the data-specific structures. They named this sample-efficient algorithm CorNet. In this paper, we consider the quality risks of RCT in complex real-world scenarios, while the above methods all use RCT as the gold standard.
\section{Empirical Method}
\subsection{Motivation and Intuition}
Given the high cost and limited availability of RCTs, increasing data volume is a practical strategy. Our setting involves 33 treatment groups with varying discount values and slight distributional differences. Directly matching each RCT group to observational data introduces bias at the group level. In contrast, the overall RCT distribution better captures the target population and reflects trade-offs across interventions. Therefore, we represent the high-dimensional data distribution using the mean and apply a bucketing strategy for fine-grained control of key features. Since inter-group differences are small, we introduce a subtle delta adjustment per sample to align intervention groups with minimal distortion. The following section shows details about how our method is conducted.

\subsection{Method Description}\label{method_des}

In the online ride-hailing pricing scenario, our treatments are 33 discrete discount types: $T = \{T_i\}_{i=1}^{N_t}$, where $N_t=33$. Under a specific discount $T_i$, the user's price will be $P*T_i \in R$ against the original price $P \in R$. After the user gets the discounted price, they may choose to take an online ride-hailing service. Therefore, the outcome is a binary variable $Y \in \{1,0\}$ to mark whether the online ride-hailing order is successfully issued. If the order is successfully issued, $Y=1$, and if it is not issued, $Y=0$. In the above process, the enterprise will comply with the user privacy agreement and record a series of desensitized features, such as the time of the current order, the supply and demand relationship of nearby vehicles, and some statistical features. These variables together constitute the high-dimensional feature $X$. 
\begin{figure}[h]
    \centering
    \includegraphics[width=1\linewidth]{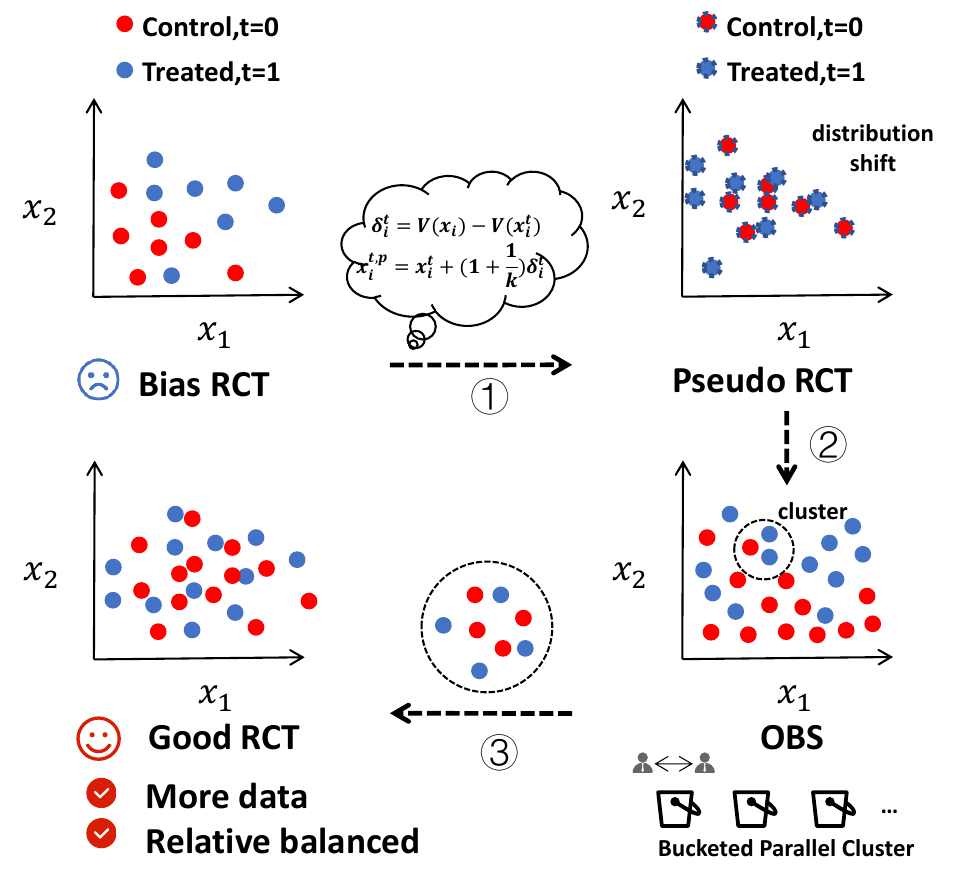}
    \caption{(1) Construct pseudo-samples based on the fusion ratio. (2) Cluster pseudo-samples from the observational data. (3) Add selected observational data back into biased RCT data to obtain good RCT.}
    \label{fig:method}
\end{figure}

To obtain the impact of discounts on order issuance, we use the uplift model to estimate the heterogeneous causal effect of $T_i$ on $Y$ of $x$ compared to $T_0$, that is,
$$ E(Y|T = T_i,X=x) - E(Y| T=T_0,X=x).$$
However, due to the small amount of RCT data, insufficient generalization, and the inherent heterogeneity of RCT data, this leads to biased results. That means if $f$ is a probability density function, then the probability density function of $X$ is different for different $T$.
$$ \exists \, T_i, T_j \in T \quad \text{let} \quad f(X \mid T = T_i) \neq f(X \mid T = T_j).$$

In our scenario, the RCT data $R = \{(T_i,X_i,Y_i)\}_{i=0}^{N_r}$ is about 0.65\% of the observational data $O=\{(T_i,X_i,Y_i)\}_{i=0}^{N_o}$. To constrain the heterogeneity between different treatments and expand the sample size, we will take a part of the observational data as a supplement to the RCT data. Specifically, the overall distribution of RCT data is closer to the target distribution than an individual RCT group. Since our data is high-dimensional and the time complexity of the algorithm needs to be considered in the industry, we calculate the feature mean of the selected $j$ partially important features $X_s \subseteq X$ dimensions for the RCT data to obtain the mean vector $V_{avg}$ of $N_r$ samples: 
\begin{equation}
\label{eq:Positional Encoding}
\begin{split}
    V_{avg} &=(\frac{1}{N_r} \sum{X_1},\frac{1}{N_r} \sum{X_2},\dots,\frac{1}{N_r} \sum{X_j}) 
    \\
 &= (\bar{X_1},\bar{X_2},\dots,\bar{X_j}).\nonumber 
\end{split}
\end{equation}
For every RCT treatment group, we obtain separate mean vectors $V$:
$$ V = \{V_1,V_2,\dots,V_{N_t}\}.$$
In data fusion, for each treatment group of RCT, we first calculate the deviation $\delta$ between its feature mean vector and the overall distribution mean vector:
\begin{equation}
\label{eq:Positional Encoding}
\begin{split}
   \delta_i &= V_{avg}-V_i
   \\ &=(\bar{X_1}-\bar{X}_{1,t=i},\dots,\bar{X_j}-\bar{X}_{j,t=i}).\nonumber 
\end{split}
\end{equation}
So we get a set of bias vectors $\delta$:
$$\delta = \{\delta_1,\dots,\delta_{N_t}\}.$$
According to the preset fusion ratio k, to make the fused mean of the treatment equal to the overall distribution mean, the data mean we need to extract from the observational data of treatment $i$ is:
$$V_i^{obs} = V_i + (1+\frac{1}{k})*\delta_i.$$
Therefore, for each sample in the RCT, we construct a pseudo sample to be fused with the target, that is
\begin{equation}
\label{eq:Positional Encoding}
\begin{split}
    R_i^{pesudo} = \{(T_i,[X_{s_i} + (1+\frac{1}{k})*\delta_i; X/X_{s_i}],Y_i)\}.
   \nonumber 
\end{split}
\end{equation}
Then we perform column normalization on the pseudo sample and the observational data so that different dimensions are scaled to the same range to avoid the interference of numerical values in the selection of observational samples. After normalization, a corresponding k data point with the closest feature to the pseudo sample is extracted from the observational data. We use the bucket first and the L2 distance:
\begin{equation}
\label{eq:Positional_Encoding}
\begin{split}
    d_{m,n} = \left\{
\begin{array}{ccl}
\infty & \text{if} & \text{bucket}_m \neq \text{bucket}_n \\
||X_s^m - X_s^n||_2 & \text{if} & \text{bucket}_m = \text{bucket}_n
\end{array} .\right.
\nonumber 
\end{split}
\end{equation}
In industrial scenarios, several important variables have a huge impact on the final effect of the model. For such variables, we hope to ensure that the heterogeneity of the features in fused data is aligned well between different treatments. Therefore, we will first bucket the observational data and pseudo-sample data according to these variables, and then calculate the similarity of the observational samples and pseudo-samples in the same bucket. In addition, the bucketing operation can make full use of the cluster environment, allowing each bucket to be calculated on a distributed machine, maximizing the concurrency of the algorithm and optimizing the algorithm time under large-scale data.  According to the different importance of features and business needs, calculating L2 in the distance measurement can add corresponding weights to different features, so that the fusion process pays more attention to the heterogeneity of these important features:$$ \sum_{k=1}^{j}{w_k*||X_{s_k}^m - X_{s_k}^n||_2} .$$
Through the above process, we can implement the data fusion algorithm at any fusion ratio k, select appropriate features in high-dimensional scenarios, and choose different distance measurement methods to adapt to complex industrial scenarios. The pseudo-code of the algorithm is shown in Algorithm \ref{alg:quicksort}. The overall workflow of the method is shown in Figure \ref{fig:method}.
\vspace{-1em}
\begin{algorithm}[h]
\caption{Pseudo-Sample Matching Algorithm}
\label{alg:quicksort}
\begin{algorithmic}
    \REQUIRE RCT data $R$, Observational data $O$, Fusion ratio $k$
    \ENSURE Fused dataset $S$

    \STATE \textbf{Function} \textsc{DataFusion}$(R, O, k)$
    \STATE $V_{\text{avg}} \leftarrow (\bar{X}_1, \bar{X}_2, \dots, \bar{X}_j)$
    \STATE $S \leftarrow \emptyset$
    \FOR{$i \leftarrow 1$ \TO $N_t$} 
        \STATE $V_i \leftarrow (\bar{X}_{1,t=i}, \dots, \bar{X}_{j,t=i})$
        \STATE $\delta_i \leftarrow V_{\text{avg}} - V_i$

        \STATE $R_i^{\text{pseudo}} \leftarrow (T_i, [X_{s_i} + (1+\frac{1}{k})*\delta_i; X/X_{s_i}], Y_i)$

        \STATE $R_i^{\text{pseudo}}, O_i \leftarrow \text{normalization}(R_i^{\text{pseudo}}, O_i)$

        \FOR{$q \leftarrow 1$ \TO number of samples in $R_i^{\text{pseudo}}$}
            \STATE $d \leftarrow \text{DISTANCE}(R_{i,q}^{\text{pseudo}}, O_i)$
            \STATE $S \leftarrow S \cup \text{k's nearest samples in } O_i$
        \ENDFOR
    \ENDFOR

    \STATE \textbf{return} $S$
    
\end{algorithmic}
\end{algorithm}
\vspace{-2em}
\subsection{Time Complexity Analysis}

 Assume that the cluster has enough computing units, and each computing unit can process a bucket in parallel. Given a total of $N_b$ buckets with a relatively balanced distribution of samples across buckets, where the number of RCT samples is $N_r$ and the number of observational samples is $N_o$, the expected number of RCT and observational samples in each bucket is approximately $\frac{N_r}{N_b}$ and $\frac{N_o}{N_b}$, respectively. For each sample, the time complexity of KNN~\cite{guo2003knn} clustering in the observational data is $O(\frac{N_o}{N_b})$, and the total time complexity is $O(\frac{N_o*N_r}{N_b^2})$. In actual industrial scenarios, the KNN algorithm can also be accelerated using algorithms such as KD trees~\cite{bentley1975multidimensional,friedman1977algorithm} to optimize the time complexity to $O(N_r*log N_o)$.
 
\section{Experimental Results}

In Section \ref{sec:metric}, we provided a discussion of the evaluation metrics used in our study. Section \ref{sec:sim} detailed the simulation experiments. Section \ref{sec:real} presented the model's performance evaluation on real-world data about offline and online evaluation.

\subsection{Evaluation Metric}\label{sec:metric}
In the simulation evaluation and offline evaluation, we use Qini coefficients, MAPE, and Click Over Predicted Click (COPC) to evaluate the uplift model. Since there are multiple intervention groups, we calculate the metrics for each intervention group and the control group, then take a weighted average based on the number of online users. Noted that we selected 15 groups in the offline evaluation, accounting for 92\% of the total population. Therefore, the optimal COPC  is expected to be approximately 0.92. In the online verification,  economic benefit serves as a direct performance metric. We calculate Gross Merchandise Value (GMV), Gross Profit (GP), and AA difference to estimate the overall business profit. 

\subsection{Simulation experiment}\label{sec:sim}
\subsubsection{Data Generate Process}
We use synthetic data for simulation analysis. To simulate the real scenario of online ride-hailing pricing, we need to follow three keys: 
\begin{itemize}
    \item The scale of observational data is much larger than RCT data.
    \item RCT data has certain heterogeneity.
    \item Features are relatively high-dimensional vectors compared to the data volume.
\end{itemize}

(1) \textbf{Covariates, interventions, and outcomes:} We set the feature covariates to 20 dimensions. In real scenarios, the covariates' distribution is complex. For this reason, the simulation data we generated includes binomial distribution, Poisson distribution, normal distribution, uniform distribution, bimodal distribution, and (non) linear combinations of the above distributions, thereby simulating complex real-world scenarios. Details are in the code repository.

 According to the covariate $X$ and the binary intervention discount $T$, we generate heterogeneous causal effects and obtain the binary label $Y$.
\begin{equation}
\begin{split}
Y &= \left\{
\begin{array}{lcl}
1& & f_y(X,T) + \delta > \mu\\
0& & f_y(X,T) + \delta < \mu\\
\end{array}\right.
   \nonumber 
\end{split}
\end{equation}
$$   \delta \sim N(0,1), $$
where
$$f_y(X, T) = a + b \cdot T + c \cdot X \cdot T + d \cdot X^2 \cdot T + e \cdot X + g \cdot X^2. $$

and $a,b,c,d,e,g,\mu$ are hyperparameters. Further, we adjusted the hyperparameters to ensure that the proportions of the four categories --- Persuadables, Sleeping Dogs, Lost Causes, and Sure Things~\cite{lo2002true} --- are reasonable and similar to the proportions observed in real-world scenarios. For example, when the amount of data is 1000, the ratio is shown in Table \ref{tab:bili}.
\begin{table}[h]
  \caption{The proportions of the four categories: Persuadables, Sleeping Dogs, Lost Causes, and Sure Things. OBS represents observational data.}\label{tab:bili}
  \begin{tabular}{c|cccc}
    \toprule
     \textbf{Categories}& $Y(T=1)$ & $Y(T=0)$ & \#RCT & \#OBS \\
    \midrule
     Persuadables &1 & 0 & 237& 23455\\
     Sure Things &1 & 1 & 183& 18683\\
     Lost Causes &0 & 0 & 560& 56083\\
    Sleeping Dogs &0 & 1 & 20& 1779\\
    \bottomrule
  \end{tabular}
\end{table}

(2) \textbf{Generation of biased RCT and observational data:} We use selection bias as the factor that causes heterogeneity in RCT data. Specifically, for different covariates X, the probability of being subjected to different interventions is different, that is,
$P(T|X) = f_b(X)$ and $\exists\ x_i,x_j ,  f_b(x_i) \neq f_b(x_j)$.

Similarly, we use selection bias as the key factor that causes heterogeneity in observational data. So
$P(T|X^{obs}) = f_o(X^{obs})$ and $\exists\ x_i^{obs},x_j^{obs} ,  f_o(x^{obs}_i) \neq f_o(x^{obs}_j) $.

Note that the difference between the above two lies in the impact of selection bias on data heterogeneity. In observational data, selective bias has a greater impact, while this difference is relatively small in biased RCT data. We adjust hyperparameters to make the heterogeneity of the observational data larger than the biased RCT data, and make the amount of observational data 100 times that of the RCT data.

(3) \textbf{Generate the Ground-Truth:} To effectively assess the true performance of our model, we construct a large-scale, unbiased RCT, 100 times larger than the biased RCT, to serve as the ground truth and gold standard for evaluating the model's true capabilities. In real-world scenarios, it is common to encounter discrepancies where offline evaluation metrics and online performance diverge. This inconsistency may stem from biases present in the RCT data. Unfortunately, in practice, obtaining a truly unbiased RCT for model evaluation is often infeasible. However, with synthetic data, we can simulate an unbiased RCT by making assumptions about the data distribution and the functional relationships among $Y$, $X$, and $T$. By comparing test results using both unbiased and biased RCTs, we can evaluate how imperfections in RCT data quality impact model performance.

\subsubsection{Experimental Setting}

We use TARNet (TAR)~\cite{shalit2017estimating} and Causal Forests(CF) \cite{wager2018estimation}, which are widely used in industry scenarios as the model for causal effect estimation. We split the generated biased RCT into a training set and a test set in an 8:2 ratio. The data volume of biased RCT is 1000, 5000, and 10000. Correspondingly, the observational data is also 100 times larger than the biased RCT.  In addition, we set the fusion ratio to 1, 3, and 5. We use the KD tree to reduce the time complexity to $O(log N)$. We evaluate ten times and report the mean and standard deviation. The 1:3 fusion ratio is considered the main experiment, and three different fusion ratios (1:1, 1:3, and 1:5) are used to explore the impact of this hyperparameter. 

\begin{table*}[h]
\caption{Our method (\_f) is compared with the baseline and a control group (\_r) where added data is randomly selected in equal amounts from observational data.}\label{tab:fangzhen}
\centering
{\small 
\begin{tabular}{c|ccc|ccc|ccc}
\hline
\multirow{2}{*}{Setting} & \multicolumn{3}{c|}{n=1000} & \multicolumn{3}{c|}{n=5000} & \multicolumn{3}{c}{n=10000} \\ \cline{2-10} 
                       & Qini & MAPE & COPC & Qini & MAPE & COPC & Qini & MAPE & COPC \\ \hline
TAR  & 0.222$\pm$0.019 & 0.273$\pm$0.089 & 0.969$\pm$0.147 & 0.259$\pm$0.007 & 0.091$\pm$0.037 & 1.047$\pm$0.090 & 0.299$\pm$0.024 & 0.170$\pm$0.101 & \pmb{1.003$\pm$0.210} \\
TAR\_r  & 0.222$\pm$0.024 & 0.259$\pm$0.125 & \pmb{0.973$\pm$0.226} & 0.295$\pm$0.008 & 0.101$\pm$0.069 & 0.950$\pm$0.923 & 0.327$\pm$0.011 & 0.108$\pm$0.089 & 0.992$\pm$0.125 \\ \hline
\textbf{TAR\_f(ours)}  & \pmb{0.250$\pm$0.027} & \pmb{0.219$\pm$0.160} & 0.927$\pm$0.150 & \pmb{0.323$\pm$0.005} & \pmb{0.078$\pm$0.074} & \pmb{0.981$\pm$0.101} & \pmb{0.341$\pm$0.004} & \pmb{0.057$\pm$0.046} & 1.021$\pm$0.073 \\ \hline
\hline
CF & 0.273$\pm$0.001 & 0.140$\pm$0.001 & \pmb{1.032$\pm$0.032} & 0.304$\pm$0.001 & 0.115$\pm$0.008 & 1.131$\pm$0.009 & 0.318$\pm$0.000 & 0.062$\pm$0.003 & 1.056$\pm$0.002 \\
CF\_r & 0.296$\pm$0.005 & 0.120$\pm$0.001 & 1.109$\pm$0.000 & 0.325$\pm$0.001 & 0.070$\pm$0.000 & 1.056$\pm$0.003 & 0.328$\pm$0.000 & 0.053$\pm$0.000 & \pmb{1.053$\pm$0.000} \\ \hline
\textbf{CF\_f(ours)} & \pmb{0.316$\pm$0.003} & \pmb{0.102$\pm$0.016} & 1.117$\pm$0.034 & \pmb{0.330$\pm$0.000} & \pmb{0.047$\pm$0.003} & \pmb{1.050$\pm$0.004} & \pmb{0.334$\pm$0.000} & \pmb{0.052$\pm$0.004} & 1.057$\pm$0.004 \\

\bottomrule
\end{tabular}

}

\end{table*}

\subsubsection{Experimental Results}

(1) \textbf{Visualization of data heterogeneity.}

\begin{figure}[h]
    \centering
    \includegraphics[width=0.9\linewidth]{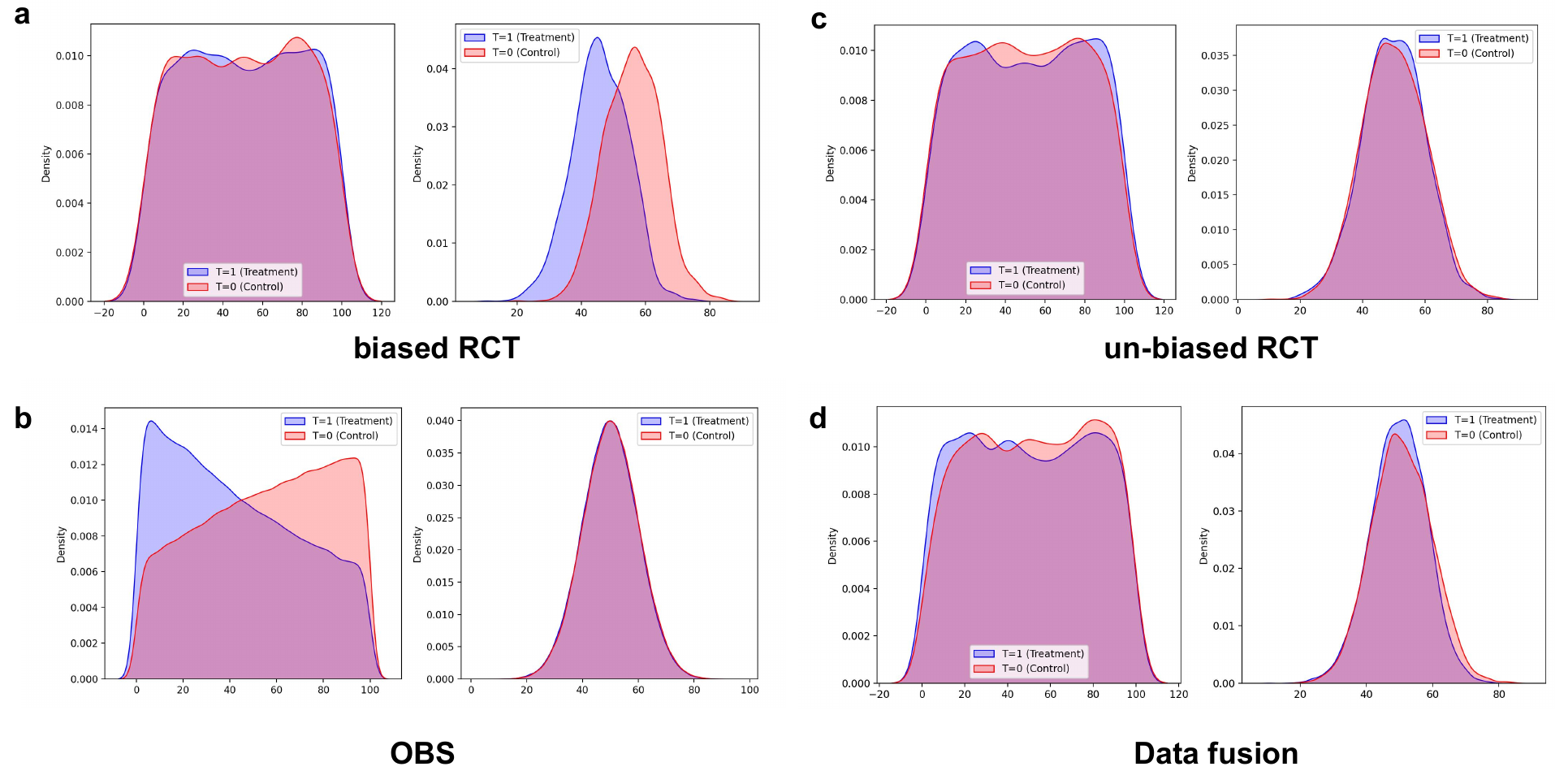}
    \caption{Illustration of data feature heterogeneity. We selected two features. (a) represents the biased RCT data, (b) represents the observational data, (c) is the generated unbiased RCT data at the same scale, and (d) represents the fused data after a 1:3 fusion.}
    \label{fig:yzxduibi}
    \vspace{-2em}
\end{figure}

We select two features with a data volume of 1000 and a fusion ratio of 3 as an example, shown in Figure \ref{fig:yzxduibi}. 

From the analysis chart, we can observe that: (i) When the biased RCT data has no heterogeneity in a certain dimension and large heterogeneity in the observational data, our method can ensure that the fused data will not introduce large deviations due to the differences in the observational data. (ii) When the biased RCT has large heterogeneity in a certain dimension, while the observational data has no heterogeneity in this dimension, our method can effectively correct the bias of the biased RCT data and ensure that the fused data does not have large heterogeneity.

(2) \textbf{Results and analysis.} We use the model with a 1:3 fusion ratio as our main experimental setting, with the baseline trained on the original biased RCT data. The control group consists of data randomly selected in a 1:3 ratio from the observational data. Then tested on a large-scale unbiased RCT as the ground truth. The results are shown in Table \ref{tab:fangzhen}. 

\begin{figure*}[h]
    \centering
    \includegraphics[width=0.8\linewidth]{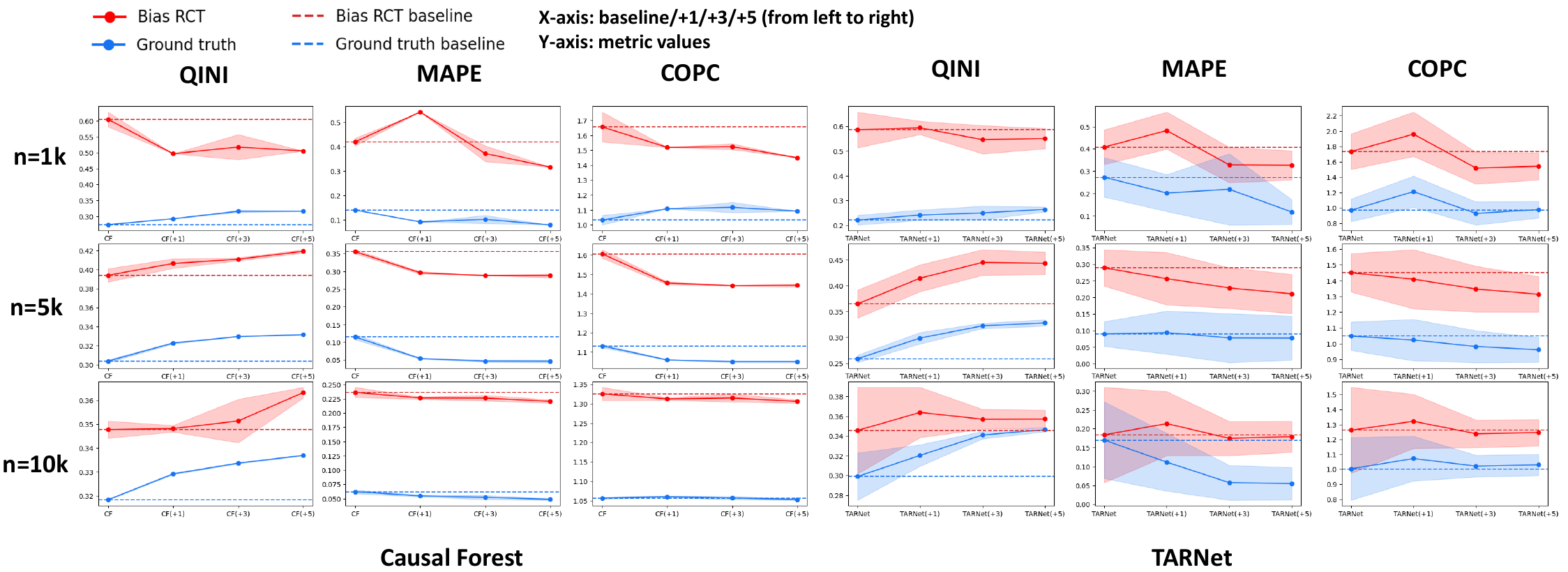}
    \caption{The variation of the Qini, Mape, and COPC under different fusion ratios and using different test sets, where \( n \) represents the data size and (+ratio) denotes the fusion ratio. In each subplot, the fusion ratio increases sequentially from left to right, with red representing the biased RCT used as the test set and blue representing the ground truth.}
    \label{fig:qini}
    \vspace{-1em}
\end{figure*}

Moreover, we tested the model trained under different fusion ratios and explored the impact when the RCT data used in industry is biased. The test results are detailed in Figure \ref{fig:qini}.

(i) Data fusion can significantly improve the model's effect on the ground truth. As shown in Table \ref{tab:fangzhen}, the evaluation metrics outperform the baseline across different data scales.

(ii) Biased and unbiased RCTs yield contradictory trends. As shown in Figure \ref{fig:qini}, increasing the fusion ratio leads to improved Qini scores under the ground truth, while performance based on biased RCTs fluctuates and deteriorates, exhibiting high variance. MAPE and COPC show the same trend. This divergence underscores the risks of relying on low-quality RCTs, which may produce misleading or even opposite conclusions, emphasizing the critical need for high-quality RCT data in industrial applications.

(iii) Fusion ratio positively influences performance. In the synthetic setting, higher fusion ratios consistently enhance Qini and reduce MAPE.

\subsection{Real World Data Verification}\label{sec:real}
\begin{figure}[h!]
    \centering
        \includegraphics[width=1\linewidth]{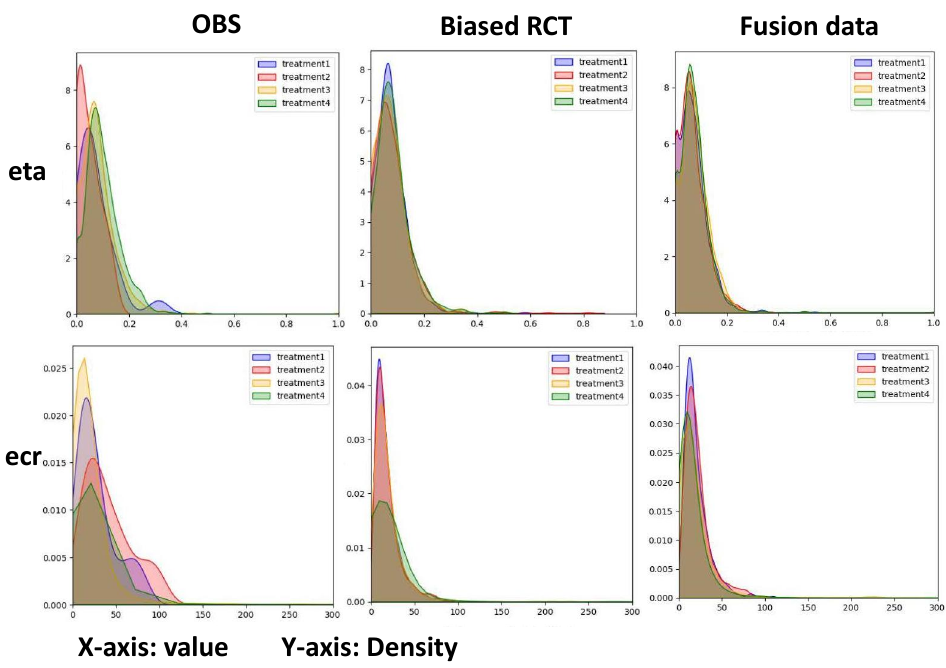}
    \caption{Visualization of heterogeneity in real high-dimensional data. We selected two features. From left to right are observation data, biased RCT data, and our fusion data.}
    \label{fig:rhyzx}
    \vspace{-2em}
\end{figure}
To test the effect of our method in real-world scenarios, we conducted the offline and online verification.  Specifically, in the online ride-hailing pricing scenario, the RCT experiment accounts for 0.65\% of the total online traffic. To train the uplift model, we use RCT data from more than 200 main cities in China collected over one year. And we use the RCT data closest to the current week as the test set. We selected the 1:3 model, which performed well in offline experiments, for online deployment. It was used for one week across 62 major cities in China, during which economic gains were observed.

\subsubsection{Offline Evaluation}

We conducted data fusion with multiple ratios using the (biased) RCT data points obtained from experiments in the aforementioned 200 cities. The fusion ratios were set to 1, 2, 3, 5, 7, and 9, respectively. During the matching process, we imposed a constraint that required the distance to RCT samples to reach at least a certain threshold. The experimental results are shown in the Table \ref{tab:real_off}.

\begin{table}[h]
\caption{The baseline represents the model trained using the original biased RCT. The ratios from 1:1 to 1:9 indicate different fusion ratios. The $(w\_)$ represents the weighted average of the 15 treatments mentioned above based on their population proportions (thus, the theoretical optimal value of w\_COPC should be close to 0.92). The gray row is what we chose in the online experiment. }\label{tab:real_off}
\begin{tabular}{c|ccc}
    \toprule
     Setting & w\_Qini & w\_MAPE & w\_COPC \\
    \midrule
    baseline & \underline{0.0907} & 0.0441 & 0.9509 \\ \hline
   
    1:1 & 0.0697 & 0.0402 & 0.8385 \\ 
    1:2 & \textbf{0.0967} & 0.0627 & \underline{0.9496}\\ 
    \cellcolor{gray!20}1:3 & \cellcolor{gray!20}0.0796 &\cellcolor{gray!20} \textbf{0.0308} & \cellcolor{gray!20}\textbf{0.9343} \\ 
    1:5 & 0.0865 & \underline{0.0360} & 0.8718 \\ 
    1:7 & 0.0891 & 0.1064 & 0.7588 \\
    1:9 & 0.0788 & 0.1396 & 1.2216 \\ 
    \bottomrule
  \end{tabular}
\vspace{-1em}
\end{table}

Upon examining the result, we observe that data fusion has produced a negative effect on the Qini coefficient across almost all experimental groups, while it has shown a positive effect on MAPE and COPC at certain fusion ratios. Since our test set is derived from biased RCT data, the results may differ from or even contradict the model's true capabilities, as indicated by the simulation experiment results in Section \ref{sec:sim}, but it still serves as an important reference for mitigating online risks. The 1:3 setting requires fewer fused data while simultaneously achieving better COPC and MAPE scores. Therefore, through further consideration of the model and cost estimation, we have selected the 1:3 model as the deployment model and conducted online experiments in the next section. In addition, we selected two key features and visualized them, with the results presented in Figure \ref{fig:rhyzx}. The eta and ecr are crucial for the supply and demand relationship, which plays a significant role in our scenario. As shown in the visualization, the fused data helps mitigate the bias and heterogeneity in the RCT data, resulting in an improved representation of the underlying relationships.

\subsubsection{Real World Online Evaluation}

Offline evaluation fails to reflect the actual online economic benefits. Therefore, we conducted one week\footnote{In the ride-hailing scenario, online performance exhibits weekly periodicity, and the large volume of online data ensures statistical significance;} of online experiments in 62 major cities in China, with the experimental traffic accounting for 20\% of the total traffic. The model trained using the 1:3 fused data was tested to verify whether it could have positive results in an online setting. The model is used to predict users' real-time sensitivity to discounts when hailing a ride online, and the predicted values serve as the basis for the coupon issuance decisions made by the budgeting system. We compared the performance of the fusion model and the baseline model regarding GMV and GP, and calculated the AA difference for the two experimental groups, as shown in Table \ref{tab:online}. The model achieved a 0.87\% improvement in GMV. Because in large-scale online tests, it is difficult to fully control all variables. Therefore, we rely on the computed adjusted profit as the primary evaluation criterion, considering the GP loss and AA difference, with the baseline as the reference. The profit gain brought by the model is +0.41\% on the overall platform, demonstrating the effectiveness of the method. 

\begin{table}[h]
  \caption{The online performance of the model trained on the original data and the 1:3 fused data. We compute and care more about the profit, which corresponds to the gray area.}\label{tab:online}
  \begin{tabular}{c|ccc|>{\columncolor{gray!20}}c}
    \toprule
     Setting & GMV & GP & AA & profit\\
    \midrule
     \textbf{fused data(ours)} & +0.87\% & -0.19& +0.15\% &\textbf{+0.41\%} \\
     baseline & 0.00\% & +0.01 & +0.09\%&0.00\% \\
    \bottomrule
  \end{tabular}
\end{table}
\section{Conclusion}

To address the issues of potential complex heterogeneity and significant generalization bias due to insufficient sample sizes in industrial RCTs, we propose a method for combining RCTs and observational data based on pseudo-sample matching. Our method leverages flexible bucketing and fusion rates to adapt to complex business scenarios. We validate its effectiveness through simulations, offline evaluation, and online experiments. In the simulation experiments, we revealed the harm caused by biased RCTs to model training and evaluation, as well as their impact on economic benefits. Our method performed well on the simulation dataset. Through offline validation with large-scale real-world data and online deployment experiments of the model in multiple major cities in China, our method achieved a 0.41\% improvement in economic profit. We introduce a new setting compared to previous studies, aiming to solve real-world industrial problems, and hope that this work provides valuable insights for handling data in industrial applications and emphasizes the critical importance of RCT quality for the uplift model performance. 

\section*{Acknowledgment} 
This work was supported in part by the National Key Research and Development Program of China (2024YFE0203700), National Natural Science Foundation of China (62376243), "Pioneer" and "Leading Goose" R\&D Program of Zhejiang (2025C02037), and the Starry Night Science Fund of Zhejiang University Shanghai Institute for Advanced Study (SN-ZJU-SIAS-0010). All opinions in this paper are those of the authors and don't necessarily reflect the views of the funding agencies.

\section*{GenAI Usage Disclosure} 
We used generative AI, ChatGPT, to check for syntactic and grammatical errors in the manuscript. We carefully verified the correctness of the revised content.
\bibliographystyle{ACM-Reference-Format}
\balance
\bibliography{sample-base}

\end{document}